\documentclass[12pt,a4paper]{article}
\usepackage{amsmath}
\usepackage{amssymb,amsfonts}
\usepackage[square,comma]{natbib}
\usepackage{latexsym}

\begin{document}
\title{\textsf{A new algorithm for anisotropic solutions}}
\author{M. Chaisi\thanks{Permanent address: Department of
Mathematics \& Computer Science, National University of Lesotho,
Roma 180, Lesotho; eMail: \texttt{m.chaisi@nul.ls}}  \; and  S. D.
Maharaj\thanks{Author for correspondence; email:
\texttt{maharaj@ukzn.ac.za}; fax: +2731
260 2632}\\
Astrophysics and Cosmology Research Unit\\
School of Mathematical Sciences \\ University of KwaZulu-Natal\\
Durban 4041, South Africa}
\date{}
\maketitle
\begin{abstract}
We establish a new algorithm that generates a new solution to the
Einstein field equations, with an anisotropic matter distribution,
from a seed isotropic solution. The new solution is expressed in
terms of integrals of an isotropic gravitational potential; and the
integration can be completed exactly for particular isotropic seed
metrics. A good feature of our approach is that the anisotropic
solutions necessarily have an isotropic limit. We find two examples
of anisotropic solutions which generalise the isothermal sphere and
the Schwarzschild interior sphere. Both examples are expressed in
closed form involving elementary functions only.
\end{abstract}

\section{Introduction \label{sec:intro}}

Numerous models of static perfect fluid spheres, in the context of
general relativity, have been constructed in the past because these
are first approximations in building a realistic model for a star.
Lists of exact solutions to the Einstein field equations modelling
relativistic perfect fluid spheres are given in several treatments
\cite{DelgatyLake,FinchSkea,stephaniEtAl}. In these works it is
assumed that the matter distribution is isotropic so that the radial
pressure is the same as the transverse pressure. A strong case can
be made to study matter distributions which are anisotropic in which
the radial component of the pressure is not the same as the
transverse component. Anisotropy should not be neglected when
analysing the critical mass and the redshift of highly compact
bodies, and it is important in modelling boson stars and strange
stars. Consequently anisotropy has been studied extensively in
recent years by a number of researchers within the framework of
general relativity \cite{DevGleiser,DevGleiser2,HerreraMartin,%
HerreraTroconis,Ivanov,MakHarko2002,MakHarko,SharmaMukherjee2002}.

Most solutions of the Einstein field equations with isotropic matter
have been obtained in an ad hoc approach. Recent investigations have
attempted to generate isotropic models using a systematic and
algorithmic
approach\cite{RahmanVisser,Lake,MartinVisser,BoonsermEtAl}. However
there exist very few analogous results for generating anisotropic
models. In this regard Maharaj and Chaisi\cite{MaharajChaisi} have
established an algorithm that produces a new anisotropic solution
from a given seed isotropic line element. Here we present  a
different algorithm that generates a new anisotropic solution. The
present algorithm has a simpler form and involves fewer
integrations, and it is consequently easier to apply. A desirable
physical feature of our approach is that these solutions have an
isotropic limit; we expect that gravity acts  to eventually
isotropize matter in the absence of other external forces. Note that
many of the exact solutions found previously remain anisotropic with
no possibility of regaining an isotropic matter distribution in a
suitable
limit\cite{MaharajMaartens,gokhroo,ChaisiMaharajA,ChaisiMaharajB}.

The main objective of this paper is to demonstrate that it is
possible to generate anisotropic solutions from a given isotropic
solution. To achieve this we need to complete an integration which
we show is possible for particular metrics. In \S2 we provide the
fundamental field equations for isotropic and anisotropic matter.
The field equations are presented as a first order system of
differential equations. The algorithm that produces a new
anisotropic solution is described in \S3. As a first example we use
the isothermal model to produce a new anisotropic solution in \S4.
As a second example we use the Schwarzschild interior model to
produce a new anisotropic solution in \S5. In both examples the
solution can be given explicitly in terms of elementary functions
which makes it possible to study the physical features. We briefly
study the behaviour of the anisotropy factor in both examples.
%
\section{Field equations \label{sec:fieldeqns}}

We utilise a form of the Einstein field equations in which only
first order derivatives appear. This representation assists in
simplifying the integration process as pointed out by Chaisi and
Maharaj\cite{ChaisiMaharajA} whose notation and conventions we
follow. The line element for static spherically symmetric spacetimes
is given by
\begin{eqnarray}
\mbox{d}s^2 & = &
-e^{\nu}\mbox{d}t^2+e^{\lambda}\mbox{d}r^2+r^2\left(\mbox{d}\theta^2
+\sin^2\theta\mbox{d}\phi^2\right) \label{metric2}
\end{eqnarray}
where $\nu(r)$ and $\lambda(r)$ are arbitrary functions. The
energy-momentum tensor, for nonradiating matter, for isotropic
distributions has the form
\begin{eqnarray}
T^{ab}=(\mu+p) u^au^b+pg^{ab} \label{Tiso}
\end{eqnarray}
where $\mu$ is the energy density and $p$ is the isotropic pressure.
These are measured relative to the comoving four-velocity
$u^a=e^{-\nu/2}\delta^a_0 $. We define the mass function as
\begin{eqnarray}
m(r) & = & \frac{1}{2}\int^r_0x^2\mu(x)\mbox{d}x \label{massFun}
\end{eqnarray}
Consequently $M=m(R)$ is the total mass of a sphere of radius $R$.
The Einstein field equations are equivalent to the system
\begin{eqnarray}
e^{-\lambda} & = & 1-\frac{2m}{r} \label{EFEs:1} \\
r(r-2m)\nu^\prime & = & p r^3+2m \label{EFEs:2} \\
\left(\mu+p\right)\nu^\prime+2p^\prime & = & 0 \label{EFEs:3}
\end{eqnarray}
where we have used (\ref{metric2})-(\ref{massFun}).

The energy-momentum tensor for anisotropic matter which is not
radiating has the form
\begin{eqnarray}
T^{ab}=(\mu+p) u^au^b+pg^{ab}+\pi^{ab} \label{Taniso}
\end{eqnarray}
The quantity
$\pi^{ab}=\sqrt{3}S(r)\left(c^ac^b-\frac{1}{3}h^{ab}\right) $ is the
anisotropic stress tensor; the spacelike vector $c^a =
e^{-\lambda/2}\delta^a_1$ is orthogonal to the fluid four-velocity
$u^a=e^{-\nu/2}\delta^a_0 $ and $|S(r)|$ is the magnitude of the
stress tensor. The Einstein field equations, with the metric
(\ref{metric2}) and the matter content (\ref{Taniso}), can be
written in the form
\begin{eqnarray}
e^{-\lambda} & = & 1-\frac{2m}{r} \label{EFEs2:1} \\
r(r-2m)\nu^\prime & = & p_r r^3+2m \label{EFEs2:2} \\
\left(\mu+p_r\right)\nu^\prime+2p^\prime_r & = &
-\frac{4}{r}\left(p_r-p_\perp\right) \label{EFEs2:3}
\end{eqnarray}
for anisotropic matter distributions. The radial pressure $p_r$ is
distinct from the tangential pressure $p_\perp$. It is convenient to
write $p_r$ and $p_\perp$ in the form
\[p_r=p+2S/\sqrt{3}, \quad\quad p_\perp=p-S/\sqrt{3}\] where $S$
provides a measure of anisotropy. Note that for isotropic matter
$p_r=p_\perp=p$ and we regain (\ref{EFEs:1})-(\ref{EFEs:3}).
%
%
%
\section{The Algorithm \label{sec:algorithm}}

In this section we establish a procedure for generating a new
anisotropic solution of the Einstein field equations from a
specified isotropic solution. We start by considering the Einstein
field equations (\ref{EFEs:1})-(\ref{EFEs:3}) with isotropic matter
distribution. We assume that an explicit solution to
(\ref{EFEs:1})-(\ref{EFEs:3}) is known where
\begin{eqnarray}
 (\nu, \lambda, m, p) & = & (\nu_0, \lambda_0, m_0, p_0)
\label{eq:isosol}
\end{eqnarray}
and functions $\nu_0,\; \lambda_0,\; m_0\;\;\mbox{and}\;\; p_0$ are
explicitly given. Then the equations in
(\ref{EFEs:1})-(\ref{EFEs:3}) are satisfied and we can write
\begin{eqnarray}
 e^{-\lambda_0} & = & 1-\frac{2m_0}{r} \label{eq:iso1}\\
 r(r-2m_0)\nu_0^\prime & = & p_0r^3+2m_0  \label{eq:iso2}\\
\left(\frac{2m_0^\prime}{r^2}+p_0\right)\nu_0^\prime + 2p_0^\prime &
= & 0 \label{eq:iso3}
\end{eqnarray}
We next consider the Einstein field equations
(\ref{EFEs2:1})-(\ref{EFEs2:3}) with anisotropic matter distribution
and seek an explicit solution. To this end we propose the possible
solution
\begin{eqnarray}
(\nu,\lambda,m,p_r,p_\perp) & = &
\left(\nu_0,\;\lambda_0+x(r),\;m_0+y(r),\;p_0+\alpha(r),\;p_0-
\frac{\alpha(r)}{2}\right)\label{eq:anisosol}
\end{eqnarray}
where $(\nu_0,\lambda_0,m_0,p_0)$ are given by (\ref{eq:isosol}) and
$x,y\;\;\mbox{and}\;\;\alpha$ are arbitrary functions, and we have
set $\alpha=2S/\sqrt{3}$ for convenience. Then the system
(\ref{EFEs2:1})-(\ref{EFEs2:3}) becomes
\begin{eqnarray}
e^{-\left(\lambda_0+x\right)} & = & 1-\frac{2m_0+2y}{r}\label{eq:aniso2:1}\\
r(r-2m_0-2y)\nu_0^\prime  & = & p_0r^3 +\alpha
r^3+2m_0+2y \label{eq:aniso2:2}\\
\left(\frac{2m_0^\prime+2y^\prime}{r^2}+p_0+\alpha\right)\nu_0^\prime+
 2p_0^\prime+2\alpha^\prime & = &
-\frac{6}{r}\alpha \label{eq:aniso2:3}
\end{eqnarray}\label{eq:aniso2}
The systems (\ref{eq:iso1})-(\ref{eq:iso3}) and
(\ref{eq:aniso2:1})-(\ref{eq:aniso2:3}) lead to
\begin{eqnarray}
x & = & -\ln \left\{1-\frac{2y}{r}e^{\lambda_0} \right\}
\label{eq:x}\\
 y & = & -\frac{\alpha r^3}{2(1+r \nu_0^\prime)}
\label{eq:y}\\
\left(\frac{2y^\prime}{r^2}+\alpha\right)\nu_0^\prime +
2\alpha^\prime & = & -\frac{6\alpha}{r} \label{eq:alphayprime}
\end{eqnarray}
We need to integrate (\ref{eq:alphayprime}) to find the function
$\alpha$. The remaining functions $x$ and $y$ are defined in terms
of $\alpha$. Two cases arise: $\alpha=0$ and $\alpha\ne 0$. If
$\alpha=0$ then (\ref{eq:x})-(\ref{eq:alphayprime}) has the solution
\begin{eqnarray}
(\alpha,x,y) & = & (0, 0, 0)\label{eq:alphaxy0soln}
\end{eqnarray}
The trivial solution (\ref{eq:alphaxy0soln}) corresponds to the
isotropic case. Thus this algorithm regains the isotropic solution
in the appropriate limit. If $\alpha\ne 0$ then we can eliminate $y$
from (\ref{eq:alphayprime}) to get
\begin{eqnarray*}
\frac{2\nu_0^\prime}{r^2}\left\{-\frac{\alpha^\prime r^3}{2(1+r
\nu_0^\prime)} -\frac{3\alpha r^2}{2(1+r \nu_0^\prime)}+\frac{\alpha
r^3}{2}\frac{\nu_0^\prime+r \nu_0^{\prime\prime}}{(1+r
\nu_0^\prime)^2} \right\} + \alpha \nu_0^\prime +2\alpha^\prime & =
& -\frac{6\alpha}{r}
\end{eqnarray*}
This differential equation can be written as
\begin{eqnarray}
 \frac{\alpha^\prime}{\alpha} - \frac{\nu_0^\prime}{2+r
\nu_0^\prime} \left\{3- r \frac{\nu_0^\prime+r
\nu_0^{\prime\prime}}{1+r \nu_0^\prime} \right\} & = &
-\left(\nu_0^\prime+\frac{6}{r}\right)\left(\frac{1+r\nu_0^\prime}{2+r
\nu_0^\prime}\right) \label{eq:alphaprime}
\end{eqnarray}
after some simplification. On integration (\ref{eq:alphaprime})
leads to
\begin{eqnarray}
\ln\alpha & = & J_\alpha+\ln k\label{eq:Jint}
\end{eqnarray}
where $\ln k$ is a constant of integration, $k\ne 0$, and we have
set
\begin{eqnarray*}
J_\alpha & = & \int \left\{ \frac{\nu_0^\prime}{2+r
\nu_0^\prime}\left(\frac{3+2r\nu_0^\prime-r^2\nu_0^{\prime\prime}}{1+r
\nu_0^\prime} \right)
-\left(\nu_0^\prime+\frac{6}{r}\right)\left(\frac{1+r\nu_0^\prime}{2+r
\nu_0^\prime}\right) \right\}\mbox{d}r
\end{eqnarray*}
We can write (\ref{eq:Jint}) in the compact form
\begin{eqnarray}
\alpha & = & k e^{ J_\alpha}\label{eq:alphaJ}
\end{eqnarray}
Equations (\ref{eq:x}), (\ref{eq:y}) and (\ref{eq:alphaJ})
correspond to anisotropic matter.

Thus if given a known isotropic solution (\ref{eq:isosol}) we can
generate a new anisotropic solution (\ref{eq:anisosol}) where
\begin{eqnarray}
\alpha & = & k e^{J_\alpha} \label{eq:alphaxy:1} \\
x & = & -\ln \left\{1-\frac{2y}{r}e^{\lambda_0}\right\}
\label{eq:alphaxy:2}\\
y & = & -\frac{\alpha r^3}{2(1+r\nu_0^\prime)} \label{eq:alphaxy:3}
\end{eqnarray}
and the integral $J_\alpha$ is given by
\begin{eqnarray*}
J_\alpha & = & \int \left\{ \frac{\nu_0^\prime}{2+r
\nu_0^\prime}\left(\frac{3+2r\nu_0^\prime-r^2\nu_0^{\prime\prime}}{1+r
\nu_0^\prime} \right)
-\left(\nu_0^\prime+\frac{6}{r}\right)\left(\frac{1+r\nu_0^\prime}{2+r
\nu_0^\prime}\right) \right\}\mbox{d}r
\end{eqnarray*}
 The integration in $J_\alpha$ can be explicitly performed as $\nu_0$ is specified
 in the isotropic solution
 (\ref{eq:isosol}). Note that (\ref{eq:alphaxy:1})-(\ref{eq:alphaxy:3})
 applies to both cases $\alpha= 0$
 and $\alpha\ne 0$. If $\alpha= 0$ we can set $k=0$ and regain the
 isotropic result (\ref{eq:alphaxy0soln}). When $\alpha\ne 0$ then
 $k\ne 0$ and we regain the anisotropic equations (\ref{eq:x}), (\ref{eq:y}) and
 (\ref{eq:alphaJ}).

It is remarkable that our simple ansatz leads to a new anisotropic
solution of the Einstein field equations. This is subject to
completing the integration in $J_\alpha$; clearly this is possible
for particular choices of the isotropic function $\nu_0$. We
demonstrate two examples of anisotropic solutions for familiar
choices of $\nu_0$ in the next two sections. The algorithm that we
have generated in this paper is easy to apply as there is only a
single integration to be performed unlike the earlier algorithm of
Maharaj and Chaisi\cite{MaharajChaisi} which is more complicated and
involves further integrations. We believe that new anisotropic
solutions that arise from our procedure are likely to produce
realistic anisotropic stellar models. We emphasise that a desirable
feature of our approach is that our models contain an isotropic
limit which is often not the case in other approaches.
%
%
\section{Example 1 \label{sec:iso}}

As a first example we demonstrate the applicability of the algorithm
in \S3 by generating anisotropic isothermal spheres. The line
element for the isothermal model\cite{SaslawMaharaj} has the form
\begin{eqnarray}
\mbox{d}s^2 & = &
-r^{\frac{4c}{1+c}}\mbox{d}t^2+\left(1+\frac{4c}{\left(1+c\right)^2}\right)\mbox{d}r^2
+r^2\left(\mbox{d}\theta^2+\sin^2\theta\mbox{d}\phi^2\right)\label{eq:isothermalLE}
\end{eqnarray}
where $c$ is a constant. The corresponding isotropic functions for
(\ref{eq:isothermalLE}) are given by
\begin{eqnarray}
\left(\nu_0,\lambda_0,m_0,p_0\right) & = & \left(\frac{4c}{1+c}\ln
r,\;\ln\left\{1+\frac{4c}{1+c}\right\},
\frac{2cr}{4c+(1+c)^2},\;\frac{4c^2}{4c+(1+c)^2}\left(\frac{1}{r^2}\right)\right)
\label{eq:isothermalISOsol}
\end{eqnarray}
The energy density function has the form
\begin{eqnarray}
\mu_0 & = &
\frac{4c}{4c+(1+c)^2}\left(\frac{1}{r^2}\right)\label{eq:mu0iso}
\end{eqnarray}
Hence (\ref{eq:isothermalISOsol}) and (\ref{eq:mu0iso}) imply that
\begin{eqnarray}
p_0 & = & c\mu_0 \label{eq:p0mu0}
\end{eqnarray}
which is a linear barotropic equation of state. Isothermal spheres
with the density profile $\mu\propto r^{-2}$ and the equation of
state (\ref{eq:p0mu0}) appear in a variety of models for both
Newtonian and relativistic
stars\cite{ChaisiMaharajA,ChaisiMaharajB}. They have been studied
extensively in astrophysics as an equilibrium approximation to more
complicated systems which are close to a dynamically relaxed
state\cite{saslaw}.

With the isotropic functions (\ref{eq:isothermalISOsol}) we can
evaluate the integral $J_\alpha$ in (\ref{eq:alphaxy:1}) and we find
\[
J_\alpha =
-\int\left(\frac{3+14c+19c^2}{1+4c+3c^2}\right)\frac{\mbox{d}r}{r}
=-\frac{3+14c+19c^2}{1+4c+3c^2}\ln r +\ln k
\]
which leads to the expressions
\begin{eqnarray*}
\alpha & = & kr^{-\frac{3+14c+19c^2}{1+4c+3c^2}}\\
x & = &-\ln
\left\{1+k\frac{4c+\left(1+c^2\right)}{(1+c)(1+5c)}r^{-\frac{1+6c+13c^2}{1+4c+3c^2}}
\right\}
\\
y & = &
-\frac{k}{2}\left(\frac{1+c}{1+5c}\right)r^{-\frac{2c+10c^2}{1+4c+3c^2}}
\end{eqnarray*}
Consequently we obtain the new line element in the form
\begin{eqnarray}
\mbox{d}s^2 & = & -r^{\frac{4c}{1+c}}\mbox{d}t^2+
\left(1+\frac{4c}{(1+c)^2}\right)
\left(1+k\frac{4c+(1+c)^2}{(1+c)(1+5c)}r^{-\frac{1+6c+13c^2}{1+4c+3c^2}}
\right)^{-1}\mbox{d}r^2 \nonumber\\ & &
+r^2\left(\mbox{d}\theta^2+\sin^2\theta\mbox{d}\phi^2\right)
\label{eq:Lelement}
\end{eqnarray}
and the matter variables have the analytic representation
\begin{eqnarray}
m & = & \frac{2cr}{4c+(1+c)^2}-\frac{k}{2}\left(\frac{1+c}{1+5c}
\right)r^{-\frac{2c+10c^2}{1+4c+3c^2}}  \\
p_r & = & \frac{1}{r^2}\frac{4c^2}{4c+(1+c)^2}+ kr^{-\frac{3+14c+19c^2}{1+4c+3c^2}} \\
p_\perp & = &
\frac{1}{r^2}\frac{4c^2}{4c+(1+c)^2}-\frac{k}{2}r^{-\frac{3+14c+19c^2}{1+4c+3c^2}}
\end{eqnarray}
The isotropic isothermal sphere model (\ref{eq:isothermalLE})
produces the anisotropic isothermal sphere model (\ref{eq:Lelement})
utilizing our algorithm. With the parameter values $k=0$, we regain
the conventional isothermal sphere.

\begin{figure}[thb]
\vspace{1.8in} \includegraphics{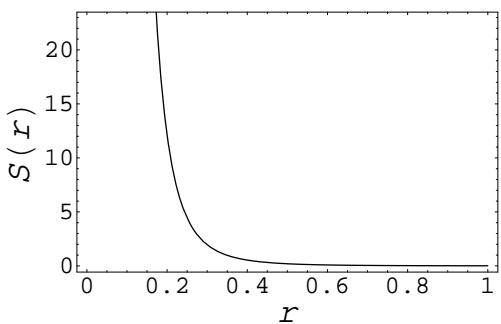} \caption{\label{fig:SisoB}Decreasing
$S(r)$ for anisotropic isothermal sphere; $c=1$, $k=.01$}
\end{figure}

\begin{figure}[thb]
\vspace{1.8in} \includegraphics{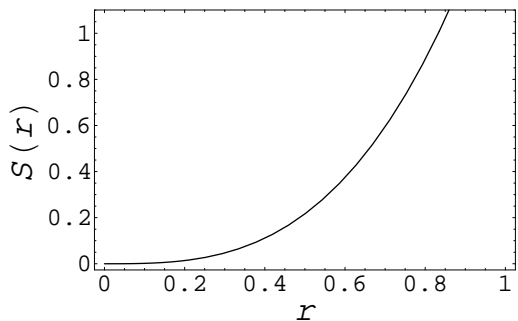} \caption{\label{fig:SisoBup}Increasing
$S(r)$ for anisotropic isothermal sphere; $c=-0.5$, $k=2$}
\end{figure}

The degree of anisotropy is
\begin{eqnarray}
S & = &
\frac{k}{2}\sqrt{3}r^{-\frac{3+14c+19c^2}{1+4c+3c^2}}\label{eq:isothermalSA}
\end{eqnarray}
Mathematica\cite{wolfram} was used to graph the anisotropy factor
(\ref{eq:isothermalSA}). The plots are as shown in Figures
\ref{fig:SisoB} and \ref{fig:SisoBup} for particular values of the
parameters shown. The anisotropy factor $S$ is plotted against the
radial distance on the interval $0< r\leq 1$. There is a singularity
at $r=0$ that has been carried over from the other dynamical and
metric functions. However because constants $k$ and $c$ can be
picked arbitrarily, the pair can be chosen such that $S(r)$ is
monotonically decreasing or increasing. The physical considerations
of a problem may lead to the choice of one profile as opposed to the
other; for example the $S(r)$ profile in Figure \ref{fig:SisoB} may
be preferable where a stellar body with vanishing anisotropy as one
moves from the center of the body to the boundary is considered. The
$S(r)$ profile in Figure \ref{fig:SisoBup} could be chosen over the
one in Figure \ref{fig:SisoB} for boson star models as proposed by
Dev and Gleiser\cite{DevGleiser}. The fairly simple behaviour of the
$S(r)$ in these plots shows that a more extensive physical analysis
of the solutions is possible, which will be carried out in future
work.

\vspace{.25in}
%
%
\section{Example 2\label{sec:Schwarz}}

As a second example we demonstrate the applicability of the
algorithm in \S\ref{sec:algorithm} by generating anisotropic
Schwarzschild spheres. The line element for the interior
Schwarzschild model\cite{DelgatyLake} is
\begin{eqnarray}
\mbox{d}s^2 & = &
-\left(A-B\Delta\right)^2\mbox{d}t^2+\Delta^{-2}\mbox{d}r^2
+r^2\left(\mbox{d}\theta^2+\sin^2\theta\mbox{d}\phi^2\right)\label{eq:schwarzsLE}
\end{eqnarray}
where $\Delta=\sqrt{1-r^2/R^2}$, $A$ and $B$ are constants. The
corresponding isotropic functions for (\ref{eq:schwarzsLE}) are
given by
\begin{eqnarray}
\left(\nu_0,\lambda_0,m_0,p_0 \right) & = &
\left(2\ln\left\{A-B\Delta\right\},\;-\ln\Delta^2,\frac{r^3}{2R^2},\;
-\frac{1}{R^2}\left[\frac{A-3B\Delta}{A-B\Delta}\right]\right)
\label{eq:schwarzsISOsols}
\end{eqnarray}
The energy density function has the form $\mu_0=3/R^2$. We therefore
have
\begin{eqnarray}
\mu_0 & = & \mbox{constant}\label{eq:schwarzsMU0}
\end{eqnarray}
for incompressible matter. This is a reasonable approximation in
particular situations as the interior of dense neutron stars and
superdense relativistic stars are of near uniform
density\cite{MaharajLeach,RhoadesRuffini}. Consequently the
assumption (\ref{eq:schwarzsMU0}) of uniform energy density is often
used to build prototypes of realistic stars in the modelling
process\cite{DevGleiser,MaharajMaartens,BowersLiang}.

The integral $J_\alpha$ in (\ref{eq:alphaxy:1}) takes the form
\begin{eqnarray}
J_\alpha & =  &\int\left\{ \left[ 2Br\left( 3-r^2\left(
\frac{2Br^2}{R^4\Delta^3\left(A+B\Delta\right)}
-\frac{2B^2r^2}{R^4\Delta^2\left(A+B\Delta\right)^2}
+\frac{2B}{R^2\Delta\left(A+B\Delta\right)}
\right) \right.\right.\right.\nonumber\\
& & \left. \left.+\frac{4Br^2}{R^2\Delta\left(A+B\Delta\right)}
\right) \right] \nonumber\\ & & \times \left[
R^2\Delta\left(A+B\Delta\right)
\left(1+\frac{2Br^2}{R^2\Delta\left(A+B\Delta\right)}\right)\left(2
+\frac{2Br^2}{R^2\Delta\left(A+B\Delta\right)}
\right) \right]^{-1} \label{eq:JalphaSchw}\\
& & \left.- \left(\frac{6}{r}
+\frac{2Br}{R^2\Delta\left(A-B\Delta\right)}\right)\left(1+\frac{2Br^2}{R^2
\Delta\left(A-B\Delta\right)}\right)\left(2+\frac{2Br^2}{R^2\Delta\left(A
-B\Delta\right)}\right)^{-1} \right\}\mbox{d}r\nonumber
\end{eqnarray}
With the substitution $u=\Delta=\sqrt{1-r^2/R^2}$,
(\ref{eq:JalphaSchw}) becomes
\begin{eqnarray*}
J_\alpha & = & \int\left(
\frac{\left(B+3Au-4Bu^2\right)\left(2B+Au-3Bu^2\right)}{\left(1-u^2\right)\left(A
-Bu\right)\left(B+Au-2Bu^2\right)}\right.
\\
& &
-\frac{3Bu^2\left(A-Bu\right)}{\left(2B+Au-3Bu^2\right)\left(B+Au-2Bu^2\right)} \\
& & \left.
+\frac{2B^2\left(1-u^2\right)\left(A-2Bu-2Au^2+3Bu^3\right)}{u\left(A
-Bu\right)\left(2B+Au-3Bu^2\right)\left(B+Au-2Bu^2\right)}
\right)\mbox{d}u
\end{eqnarray*}
The above integral can be simplified with the help of partial
fractions. We obtain
\begin{eqnarray*}
J_\alpha & = & \int \left(\frac{1}{u}+\frac{3}{2\left(1-u\right)}-
\frac{3}{2\left(1+u\right)}+\frac{B}{A-Bu}
+\frac{A-6Bu}{2B+Au-3Bu^2}\right.\\ & &
\left.-\frac{2A-7Bu}{B+Au-2Bu^2}\right)\mbox{d}u \\
& = & \ln
u-\frac{3}{2}\ln\left\{1-u\right\}-\frac{3}{2}\ln\left\{1+u\right\}-
\ln\left\{A-Bu\right\}+\ln\left\{2B+Au-3Bu^2\right\}\\
& &
-\int\left(\frac{A/B}{\frac{A^2+8B^2}{16B^2}-\left(u-\frac{A}{4B}\right)^2}
-\frac{\left(7/2\right)\left(u-\frac{A}{4B}\right)}{\frac{A^2+8B^2}{16B^2}-
\left(u-\frac{A}{4B}\right)^2}\right)\mbox{d}u\\
& = & \ln\left\{
\frac{u\left(2B+Au-3Bu^2\right)}{\left(A-Bu\right)\left(1-
u^2\right)^{\frac{3}{2}}\left(B+Au-2Bu^2\right)^{\frac{7}{4}}}\right\}\\
& & +\frac{A}{4\sqrt{A^2+8B^2}}\ln\left\{\frac{1-\frac{4B}{\sqrt{A^2
+8B^2}}\left(u-\frac{A}{4B}
\right)}{1+\frac{4B}{\sqrt{A^2+8B^2}}\left(u-\frac{A}{4B}\right)}\right\}
\end{eqnarray*}
We have completed the integration and obtained $J_\alpha$ in terms
of the intermediate variable $u$. In terms of the original variable
$r$ (in $\Delta$) we can write $J_\alpha$ as
\begin{eqnarray*}
J_\alpha & = & \ln\left\{\frac{\Delta\left(2B+
A\Delta-3B\Delta^2\right)}{\left(A-B\Delta\right)\left(1-
\Delta^2\right)^{3/2}\left(B+A\Delta-2B\Delta^2\right)^{7/4}}\right\}\\
& & +\frac{A}{4\sqrt{A^2+8B^2}}\ln\left\{\frac{1-
\frac{4B}{\sqrt{A^2+8B^2}}\left(\Delta-\frac{A}{4B}
\right)}{1+\frac{4B}{\sqrt{A^2+8B^2}}\left(\Delta-\frac{A}{4B}\right)}\right\}
\end{eqnarray*}

Then the function $\alpha$ in (\ref{eq:alphaxy:1}) becomes
\begin{eqnarray}
\alpha & = &
\frac{kR^3\Delta\left(2B+A\Delta-3B\Delta^2\right)}{r^3\left(A
-B\Delta\right)\left(B+A\Delta-2B\Delta^2\right)^{7/4}}
\left(\frac{1-\frac{4B}{\sqrt{A^2+8B^2}}\left(\Delta-\frac{A}{4B}
\right)}{1+\frac{4B}{\sqrt{A^2+8B^2}}\left(\Delta-\frac{A}{4B}
\right)}\right)^{\frac{A}{4\sqrt{A^2+8B^2}}} \label{eq:alphaSchw2}
\end{eqnarray}
and (\ref{eq:alphaxy:2}) and (\ref{eq:alphaxy:3}) respectively lead
to
\begin{eqnarray}
x & = & -\ln\left\{1+\frac{k R^3\left(2B+A\Delta-
3B\Delta^2\right)}{r\Delta\left(A-B\Delta\right)\left(B+
A\Delta-2B\Delta^2\right)^{7/4}}\left(1+
\frac{2Br^2}{R^2\Delta\left(A-B\Delta\right)}\right)^{-1} \right.\nonumber\\
& &\left.
\times\left(\frac{1-\frac{4B}{\sqrt{A^2+8B^2}}\left(\Delta-\frac{A}{4B}
\right)}{1+\frac{4B}{\sqrt{A^2+8B^2}}\left(\Delta-\frac{A}{4B}
\right)}\right)^{\frac{A}{4\sqrt{A^2+8B^2}}}
\right\}\label{eq:xySchw:1}
\end{eqnarray}
\begin{eqnarray}
y & = & -\frac{k
R^3\Delta\left(2B+A\Delta-3B\Delta^2\right)}{2\left(A-
B\Delta\right)\left(B+A\Delta-2B\Delta^2\right)^{7/4}}\left(1+
\frac{2Br^2}{R^2\Delta\left(A-B\Delta\right)}\right)^{-1}\nonumber\\
& &
\times\left(\frac{1-\frac{4B}{\sqrt{A^2+8B^2}}\left(\Delta-\frac{A}{4B}
\right)}{1+\frac{4B}{\sqrt{A^2+8B^2}}\left(\Delta-\frac{A}{4B}
\right)}\right)^{\frac{A}{4\sqrt{A^2+8B^2}}} \label{eq:xySchw:2}
\end{eqnarray}

Hence the new line element has the form
\begin{eqnarray}
\mbox{d}s^2 & = &
-\left(A-B\Delta\right)^2\mbox{d}t^2\nonumber\\
& & + \frac{1}{\Delta^2}\left[1+\frac{}{}\frac{}{}\frac{k
R^3\left(2B
+A\Delta-3B\Delta^2\right)}{r\Delta\left(A-B\Delta\right)\left(B+
A\Delta-2B\Delta^2\right)^{7/4}}\left(1+\frac{2Br^2}{R^2\Delta\left(A-
B\Delta\right)}\right)^{-1} \right.\nonumber\\
& &\left.
\times\left(\frac{1-\frac{4B}{\sqrt{A^2+8B^2}}\left(\Delta-\frac{A}{4B}
\right)}{1+\frac{4B}{\sqrt{A^2+8B^2}}\left(\Delta-\frac{A}{4B}
\right)}\right)^{\frac{A}{4\sqrt{A^2+8B^2}}} \right]^{-1}
\mbox{d}r^2
+r^2\left(\mbox{d}\theta^2+\sin^2\theta\mbox{d}\phi^2\right)\label{eq:schwarzsLE2}
\end{eqnarray}
and the matter variables have the form
\begin{eqnarray}
m & = & \frac{r^3}{2R^2} -\frac{}{}\frac{}{}\frac{k
R^3\Delta\left(2B+
A\Delta-3B\Delta^2\right)}{2\left(A-B\Delta\right)\left(B+
A\Delta-2B\Delta^2\right)^{7/4}}\left(1+
\frac{2Br^2}{R^2\Delta\left(A-B\Delta\right)}\right)^{-1}\nonumber\\
& &
\times\left(\frac{1-\frac{4B}{\sqrt{A^2+8B^2}}\left(\Delta-\frac{A}{4B}
\right)}{1+\frac{4B}{\sqrt{A^2+8B^2}}\left(\Delta-\frac{A}{4B}
\right)}\right)^{\frac{A}{4\sqrt{A^2+8B^2}}}
\end{eqnarray}
\begin{eqnarray}
 p_r & = &
-\frac{A-3B\Delta}{R^2\left(A-B\Delta\right)}+ \frac{k
R^3\Delta\left(2B+A\Delta-3B\Delta^2\right)}{r^3\left(A
-B\Delta\right)\left(B+A\Delta-2B\Delta^2\right)^{7/4}}\nonumber\\
& & \times
\left(\frac{1-\frac{4B}{\sqrt{A^2+8B^2}}\left(\Delta-\frac{A}{4B}
\right)}{1+\frac{4B}{\sqrt{A^2+8B^2}}\left(\Delta-\frac{A}{4B}
\right)}\right)^{\frac{A}{4\sqrt{A^2+8B^2}}}
\end{eqnarray}
\begin{eqnarray}
 p_\perp & = &
-\frac{A-3B\Delta}{R^2\left(A-B\Delta\right)}- \frac{}{}\frac{k
R^3\Delta\left(2B+A\Delta- 3B\Delta^2\right)}{2r^3\left(A-
B\Delta\right)\left(B+A\Delta-2B\Delta^2\right)^{7/4}}\nonumber\\
& &
\times\left(\frac{1-\frac{4B}{\sqrt{A^2+8B^2}}\left(\Delta-\frac{A}{4B}
\right)}{1+\frac{4B}{\sqrt{A^2+8B^2}}\left(\Delta-\frac{A}{4B}
\right)}\right)^{\frac{A}{4\sqrt{A^2+8B^2}}} \label{last}
\end{eqnarray}
The isotropic Schwarzschild sphere model (\ref{eq:schwarzsLE})
generates the anisotropic Schwarzschild sphere model
(\ref{eq:schwarzsLE2})-(\ref{last}). With the parameter value $k=0$
we regain the original interior Schwarzschild sphere.

\begin{figure}[thb]
\vspace{1.8in} \includegraphics{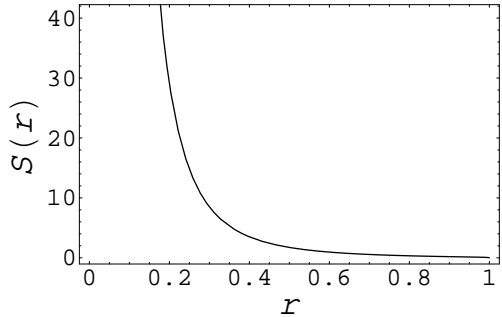} \caption{\label{fig:SchwrzIsoBup}
$S(r)$ for anisotropic Schwarzschild sphere; $A=3$, $B=1$, $k=1$ and
$R=1$.}
\end{figure}

The degree of anisotropy has the form
\begin{eqnarray}
S & = & \frac{}{}\frac{\sqrt{3}k
R^3\Delta\left(2B+A\Delta-3B\Delta^2\right)}{2r^3\left(A-
B\Delta\right)\left(B+A\Delta-
2B\Delta^2\right)^{7/4}}\left(\frac{1-
\frac{4B}{\sqrt{A^2+8B^2}}\left(\Delta-\frac{A}{4B}
\right)}{1+\frac{4B}{\sqrt{A^2+8B^2}}\left(\Delta-\frac{A}{4B}
\right)}\right)^{\frac{A}{4\sqrt{A^2+8B^2}}}\label{eq:schwarzSB}
\end{eqnarray}
Mathematica\cite{wolfram} was once again used to plot the
anisotropic factor (\ref{eq:schwarzSB}). The resulting plot is shown
in Figure \ref{fig:SchwrzIsoBup} for chosen particular values of the
parameters $A$, $B$, $k$ and $R$. Other choices of these parameters
may produce a different behaviour for $S$. The plot of $S$ against
$r$ is in the interval $0<r\leq 1$. The fact that the anisotropic
factor is in closed form and the profile as shown in Figure
\ref{fig:SchwrzIsoBup} shows that physical analysis of this model
can be investigated, which will be pursued at a later stage.

\section*{Acknowledgements}

SDM and MC thank the National Research Foundation of South Africa
for financial support. MC is grateful to the University of
KwaZulu-Natal for a scholarship.
%
%


\begin{thebibliography}{0}

\bibitem{DelgatyLake}
M S R  Delgaty and K Lake, {\it Comput. Phys. Commun.}
 {\bf 115}, 395 (1998)

\bibitem{FinchSkea}
 M R Finch and J F E Skea,
 Preprint available on the web:
  \\ http://edradour.symbcomp.uerj.br/pubs.html (1998)

\bibitem{stephaniEtAl}
 H Stephani, D Kramer, M A H  MacCullum, C Hoenslaers and
  E Herlt, {\it Exact solutions of Einstein's field equations}
  (Cambridge University Press, Cambridge, 2003)

\bibitem{DevGleiser}
 K Dev and M Gleiser,
  {\it Gen. Rel. Grav.} { \bf 34}, 1793 (2002)

\bibitem{DevGleiser2}
K  Dev and M Gleiser,  {\it Gen. Rel. Grav.} {\bf 35},
  1435 (2003)

\bibitem{HerreraMartin}
 L Herrera, J Martin and J Ospino,
  {\it J. Math. Phys.} {\bf 43},
  4889 (2002)

\bibitem{HerreraTroconis}
 L Herrera, A D Prisco, J Martin, J Ospino,
 N O Santos and O Troconis,
 {\it Phys. Rev. D} {\bf 69},
 084026 (2004)

\bibitem{Ivanov}
B V Ivanov, {\it Phys. Rev. D} {\bf 65}, 10411 (2002)

\bibitem{MakHarko2002}
M K Mak and T Harko, {\it Chin. J. Astron. Astrophys.}
  {\bf 2}, 248 (2002)

\bibitem{MakHarko}
 M Mak and T Harko,
{\it Proc. Roy. Soc. Lond. A}
  {\bf 459}, 393 (2003)

\bibitem{SharmaMukherjee2002}
 R Sharma and S Mukherjee,
 {\it Mod. Phys. Lett. A} {\bf 17},
  2535 (2002)

\bibitem{RahmanVisser}
 S Rahman and M Visser,
{\it Class. Quantum Grav.} {\bf 19}, 935 (2002)

\bibitem{Lake}
K Lake {\it Phys. Rev. D} {\bf 67}, 104015 (2003)

\bibitem{MartinVisser}
D Martin and M Visser, {\it Phys. Rev. D} {\bf 69},
  104028 (2004)

\bibitem{BoonsermEtAl} P Boonserm, M Visser and S Weinfurtner,
  ArXiv:gr-qc/0503007 (2005)

\bibitem{MaharajChaisi}
S D Maharaj and M Chaisi, {\it Math. Meth. Appl. Sci.} {\bf 29} 67
(2006)

\bibitem{MaharajMaartens}
S D Maharaj and R Maartens, {\it Gen. Rel. Grav.} {\bf 21}, 899
(1989)

\bibitem{gokhroo}
M K Gokhroo and A L Mehra, {\it Gen. Rel. Grav.} {\bf 26}, 75 (1994)

\bibitem{ChaisiMaharajA}
M Chaisi and S D Maharaj, {\it Gen. Rel. Grav.} {\bf 37}
  1177 (2005)

\bibitem{ChaisiMaharajB}
 M Chaisi and S D Maharaj, {\it Pramana - J. Phys.}
   Submitted (2006)

\bibitem{SaslawMaharaj}
 W C Saslaw, S D Maharaj, and N Dadhich,
  {\it Astrophys. J.} {\bf 471},
 571 (1996)

\bibitem{saslaw}
 W C Saslaw,
 {\it Gravitational physics of stellar and galactic systems}
 (Cambridge University Press, Cambridge, 2003)

\bibitem{wolfram}
 S Wolfram,
  {\it Mathematica}
  (Wolfram, Redwood City, 2003)

\bibitem{MaharajLeach}
S D Maharaj and P G L Leach, {\it J. Math. Phys.} {\bf 37}, 430
(1996)

\bibitem{RhoadesRuffini}
 C E Rhoades and R Ruffini,
 {\it Phys. Rev. Lett.} {\bf 32},
  324 (1974)

\bibitem{BowersLiang}
 R L Bowers and E P T Liang,
 {\it Astrophys. J.} {\bf 188}
  657 (1974)
\end{thebibliography}
\end{document}